# Method to measure neutron beam polarization with 2×1 Neutron Spin Filter

S. Masalovich

*Technische Universität München, FRM-II, D-85747 Garching, Germany*
E-mail: Sergey.Masalovich@frm2.tum.de

**Abstract**

A method to measure a beam polarization with the use of polarized $^3$He gas is discussed. It is shown that special design of the Neutron Spin Filter cell allows for a fast and accurate measurement. The accuracy of this method is analyzed.

# 1. Introduction

The progress achieved in a large-scale production of a dense spin-polarized $^3$He gas is of significant importance for many research areas where a polarized gas is considered as a subject or a tool for the investigations. In particular, it makes a great impact on the instrumentation for neutron polarization and polarization analysis since polarized nuclei of helium-3 possess very high spin-dependent neutron absorption efficiency over a wide range of neutron energy. Neutron spin filters (NSF) based on a dense hyperpolarized $^3$He gas may compete in polarization efficiency with such common devices as magnetized single crystals or supermirrors [1]. Although these latter are rather simple in operation, their applications are strongly limited by acceptable neutron energy and scattering angle range. By contrast, the broadband neutron spin filters can be sized and shaped in such a way that they will meet just about all practical needs. Until recently, the main disadvantage of this technique was a limited possibility to produce a highly polarized $^3$He gas. However, the progress achieved during last years in this area nearly eliminates this limitation. At present the gas polarization of around 50% - 70% is viable [2 - 4] through one of the optical pumping methods. In the Metastability Exchange Optical Pumping (MEOP) method direct optical pumping of metastable $^3$He atoms in $^3$He plasma at about 1 mbar is utilized [5]. After subsequent compression, the polarized gas at a pressure of a few bars may be collected in a detachable cell of a given size and shape and then transported to a neutron instrument. Another method to produce a highly polarized $^3$He gas is based on optical pumping of Rb atoms in a high-pressure Rb-$^3$He mixture [6]. The polarized Rb atoms subsequently transfer their polarization to the $^3$He nuclei in a spin-exchange process thus allowing to obtain at once a highly polarized dense $^3$He gas. This method is called Spin Exchange Optical Pumping (SEOP). Both methods are successfully employed for neutron polarization and polarization analysis and NSFs showed already a high efficiency over a wide range of experimental conditions [3, 7 - 10].

Of particular interest is the application of a polarized $^3$He gas for a very accurate measurement of a neutron polarization. For example, the present day accuracy for the measured correlation coefficients in the decay of polarized free neutrons is limited mostly by the accuracy of the measured neutron beam polarization. For further progress the uncertainty of latter below 0.1% is required [11, 12]. The methods to measure neutron beam polarization at the level of accuracy of 0.1%, which do not include the polarized $^3$He gas, have been analyzed carefully in Ref. [12]. As dense highly polarized $^3$He gas has become available, two different methods for high precision measurements of a neutron beam polarization have been proposed. In the first method an initially unpolarized neutron beam passes through a spin-polarized $^3$He gas and becomes polarized. This polarization can be evaluated with high accuracy by comparing transmission of the neutron beam through the NSF with polarized and depolarized $^3$He gas [13] (general description will be given below). This method has been tested in the neutron energy range from 40meV to 10eV and the achieved accuracy was shown to be 0.3% [14]. Nevertheless, a limitation of this method is that a precise polarization is known only when the NSF serves as a polarizing device. In many cases a neutron beam at an instrument is already polarized with some other polarizer optimized for the neutron wavelength bandwidth and the instrument geometry and the problem mostly lies in the precise measurement of this polarization. In the second method therefore the application of an "ideal" analyzer with practically 100% analyzing efficiency is proposed. This analyzer is based on the use of an opaque spin filter [11, 15], which almost completely suppresses the transmission of one spin component of the neutron beam while another spin component has noticeable, even if very low, transmission. Evidently, this method can be applicable only to the neutron beams of high intensity since the analyzing efficiency of the NSF with the gas polarization of 50 - 70% can achieve 0.999 only with the reduction of the total intensity of the transmitted beam by the factor of 100 – 10, respectively.

In the present paper we propose a method that is free of abovementioned limitations. Here a NSF cell with a special design can be used as either a polarizing or analyzing device and allows for a precise measurement of a neutron beam polarization in situ without tremendous loss of intensity.

## 2. Neutron spin filter, general description

Let us consider first a generalized spin filter [16]. In such a device the total cross section is strongly spin-dependent and for the spin-up (+) and spin-down (-) components of the neutron beam (neutron spin parallel and antiparallel to the applied magnetic field, respectively) it can be given by

$$\sigma_{\pm} = \sigma_0 \pm \sigma_p \tag{1}$$

Here $\sigma_0$ is the spin-independent part of the total cross-section and $\sigma_p = \frac{1}{2} \cdot (\sigma_+ - \sigma_-)$ is the so-called polarization cross-section. The latter is related to either spin selective scattering or spin selective absorption of neutrons in the filter material. We may now define the total transmission of an incident unpolarized neutron beam as

$$T = \frac{T_+ + T_-}{2} \tag{2}$$

and the polarizing efficiency of the filter as

$$P = \frac{T_+ - T_-}{T_+ + T_-} \tag{3}$$

Where $T_+$ and $T_-$ are the transmissions for the spin-up (+) and spin-down (-) components.

$$T_+ = \exp[-(\sigma_0 + \sigma_p)Nd] \tag{4}$$

$$T_- = \exp[-(\sigma_0 - \sigma_p)Nd] \tag{5}$$

Here N is the number of atoms per unit volume. Thus, for the filter of thickness d one can write

$$T = \exp(-\sigma_0 Nd) \cdot \cosh(\sigma_p Nd) \tag{6}$$

$$P = -\tanh(\sigma_p Nd) \tag{7}$$

It is worth noting that the equations (6) and (7) suggest no spin-flip processes occur as a neutron beam passes through the filter.

A magnetized iron plate was the first neutron spin filter and because of its simplicity it has been used extensively in scattering experiments. However, for the magnetized iron plate (6) and (7) are only applicable with limited accuracy since iron can never be completely magnetically saturated and misalignment of magnetic domains leads to a noticeable contribution of the spin-flip processes into the final polarization of the transmitted beam. A NSF based on a polarized $^3$He gas is practically free of this limitation. Penttila and Bowmann [17], for example, estimated the spin-flip probability in the windows of the NSF cell made of GE180 glass and showed it to be about $3 \times 10^{-4}$. Moreover, spin-flipped neutrons are scattered into $4\pi$ and only a very small part of them would reach a detector thus resulting in a negligible uncertainty in measuring neutron beam polarization. In addition, incoherent scattering by the $^3$He atoms can also contribute to the depolarization of a neutron beam. This contribution is controlled by the value of the incoherent cross section of $^3$He ($\sigma_{inc}$ = 0.62b for thermal neutrons [18]) and can be evaluated to be much less than $1 \times 10^{-4}$ for the $^3$He gas cell with the effective thickness $pd = 10 bar \cdot cm$. Thus, one may conclude that for a $^3$He NSF equations (6) and (7) are valid with high accuracy.

The general formulas that describe the transmittance T of a spin-filter for an incident unpolarized monochromatic neutron beam and neutron polarization P after passing through a cell with a polarized $^3$He gas can be written by analogy with (6) and (7) as follows:

$$T = T_0 \cdot e^{-\eta} \cdot \cosh(\eta P_{He}) \qquad (8)$$
$$P = \tanh(\eta P_{He}) \qquad (9)$$

Here $T_0$ – neutron beam transmission measured for the evacuated NSF cell, $P_{He}$ – polarization of the $^3$He gas, $\eta$ - cell opacity. The opacity of the cell at room temperature can be estimated from the following relation:

$$\eta = 7.32 \cdot 10^{-2} \cdot p[bar] \cdot d[cm] \cdot \lambda[A] \qquad (10)$$

where p - gas pressure, d - neutron flight pass length in the gas and $\lambda$ - neutron wavelength. Dependence of the opacity on the neutron flight pass length brings about the general constraint on the cell shape. This constraint results from the requirement of the homogeneous polarizing or analyzing efficiency over the neutron beam size. To fulfill this requirement the cell has to be shaped either with flat parallel windows in case of parallel incident and transmitted neutron beams or with concentric cylindrical (or spherical) windows in case of wide angle scattering experiments.

It follows from (9) and (10) that for an accurate estimate of the polarizing (or analyzing) efficiency of the NSF it is necessary to know the opacity (and thus the gas pressure, flight pass length and neutron wavelength) as well as $^3$He gas polarization with the same or better accuracy. The last requirement is not easy to fulfill since $^3$He polarization is usually known with limited accuracy. Moreover, $^3$He polarization, after the cell has been detached from the pumping station, has a tendency to decrease. The relaxation time of this process can vary drastically for different cells under different experimental conditions. Thus, the problem of precise measurement of the polarizing efficiency of a NSF turns into the problem of precise measurement of $^3$He gas polarization.

The more robust method to measure a beam polarization after passing the NSF can be derived directly from (8) and (9) [19]. Indeed, if one substitutes $P_{He} = 0$ into (8) then we can write the expression for the transmittance of the cell with depolarized $^3$He gas as:

$$T_{dep} = T_0 \cdot e^{-\eta} \qquad (11)$$

Then, solving the system (8-9), we arrive at

$$P = \sqrt{1 - \frac{T_{dep}^2}{T^2}} \qquad (12)$$

Hence, measuring only transmissions of the neutron beam when passing through the cell with polarized and depolarized $^3$He gas one can derive exact values for the beam polarization after NSF. To measure the transmission with depolarized gas it was proposed to use a second cell ("reference cell") filled with an unpolarized gas. This second cell has to be either identical to the first one or some special joint calibration has to be done a priori to provide a continuous on-line calibration of the neutron polarization [13]. Obviously the requirement for the measurements with depolarized gas while running an experiment brings about noticeable complication. Moreover, if a MEOP method is used to polarize a $^3$He gas, then one may expect a NSF cell to be recharged and replaced a few times during a long running experiment and hence a previous calibration will be probably lost. This certainly may result in requirement for a new calibration. Below we propose an on-line method to measure neutron polarization, which is based entirely on the data measured with polarized gas and does not require any reference cell.

### 3. 2×1 NSF cell and unpolarized incident neutron beam

To be able to determine a neutron beam polarization using only data measured with polarized $^3$He gas we need an additional equation that would replace the equation (11) and allows together with (8) and (9) to solve the system for P. As a replacement one can propose to measure a transmission of the cell along any other path of a different length. The measurements with two different paths can be done one after another in a short time thus the gas polarization can be considered to be the same. For this purpose we propose a special cell design that provides the ability to measure transmission of a neutron beam for two flight pass lengths with length ratio 2:1 (2×1 NSF cell). The design of such a cell may be, for example, a rectangular cell with two sides sized 2:1 or a cell which depth along the neutron beam can be varied stepwise with depth ratio 2:1, or just a cell with parallel windows which can be turned in such a way that the neutron flight pass length is doubled. Hence, for initially unpolarized beam we can write a set of equations:

$$T(1) = T_0(1) \cdot e^{-\eta} \cdot \cosh(\eta P_{He}) \qquad (13)$$

$$T(2) = T_0(2) \cdot e^{-2\eta} \cdot \cosh(2\eta P_{He}) \equiv T_0 \cdot e^{-2\eta} \cdot \cosh^2(\eta P_{He}) \cdot [1 + \tanh^2(\eta P_{He})] \qquad (14)$$

$$P(1) = \tanh(\eta P_{He}) \qquad (15)$$

where arguments "1" and "2" relate to the different flight pass lengths (cell position "1" and "2", respectively). Solution to this set of equations is:

$$P(1) = \sqrt{\frac{T_0^2(1)}{T_0(2)} \cdot \frac{T(2)}{T^2(1)} - 1} \qquad (16)$$

If we assume the same thickness for all windows then (16) can be slightly simplified:

$$P(1) = \sqrt{T_0 \cdot \frac{T(2)}{T^2(1)} - 1} \qquad (17)$$

With these formulas it is possible now to get information on the neutron beam polarization by measuring merely the transmission of the incident unpolarized beam along two different lengths of the same cell. The transmittance of the empty cell $T_0$ (or the ratio $T_0^2(1)/T_0(2)$ in the case of (16)) can be measured beforehand and it depends neither on the cell opacity nor the $^3$He gas polarization values. In addition, it is worth noting that either of the two positions of a 2×1 NSF cell can be used to polarize beam on a neutron instrument. The following relation gives the polarization of the beam passed along the longer side:

$$P(2) = \frac{2P(1)}{1+P^2(1)} \qquad (18)$$

## 4. 2×1 NSF cell and polarized incident neutron beam

So far an unpolarized neutron beam incident on a NSF has been considered and the polarization of a transmitted beam has been in question. In this section the application of a 2×1 NSF cell will be extended on the problem of measuring polarization of incoming polarized neutron beam. This problem finds a great practical interest in neutron instrumentation and will be considered here in more detail. Let us assume the incident beam has a polarization $P_0$. Then for the cell with a single pass length (to be specific, let us assume the cell is in the position "1") equations (8) and (9) can be modified as follows:

$$T = T_0 \cdot e^{-\eta} \cdot \cosh(\eta P_{He}) \cdot [1 + P_0 \cdot \tanh(\eta P_{He})] \qquad (19)$$

$$P = \frac{\tanh(\eta P_{He}) + P_0}{1 + P_0 \cdot \tanh(\eta P_{He})} \qquad (20)$$

Whereas equation (20) looks unsuitable for measuring polarization of an incident beam, it shows, however, that the application of a NSF might be helpful when an increase in existing polarization of a neutron beam is required. By contrast, equation (19) gives a practical way to determine $P_0$ since it is related to measured data directly. The usual method is based on measuring a flipping ratio, that is, the ratio of the transmitted intensities measured with direct and reversed beam polarization. The reverse of the polarization can be done with the use of a proper spin-flipper device. The flipping ratio is given then by:

$$R = \frac{T}{T_F} = \frac{1 + P_0 \cdot \tanh(\eta P_{He})}{1 - F \cdot P_0 \cdot \tanh(\eta P_{He})} \qquad (21)$$

Here T is the transmittance measured with the incident polarized beam (spin-flipper is off), $T_F$ is the transmittance measured with the reversed polarization of the beam (spin-flipper is on) and F is the flipping efficiency of the spin-flipper defined via relation $P_{out} = -F \cdot P_{in}$. Solution to (21) is:

$$P_0 = \frac{1}{\tanh(\eta P_{He})} \cdot \frac{R-1}{R \cdot F + 1} \qquad (22)$$

Thus, for accurate measuring of the polarization of the incident beam it is necessary, as in the case of (9), to know the opacity of the cell and $^3$He gas polarization with the same or better accuracy.

The use of a 2×1 NSF cell helps to eliminate these problems. All one has to do in this case is to measure the transmission of a neutron beam with two different states of polarization through the cell with two different flight pass lengths. The first polarization state is initial polarization to be measured and the second polarization state can be obtained from the first one by setting either a spin-flipper or depolarizer before a NSF. Although these two approaches are in many ways similar to each other we consider the first one since a spin-flipper is usually integrated in any neutron instrument with polarization analysis. Thus, by analogy with (21), one can write the flipping ratio for the short path length:

$$R_1 \equiv \frac{T(1)}{T_F(1)} = \frac{1 + P_0 \cdot \tanh(\eta P_{He})}{1 - F \cdot P_0 \cdot \tanh(\eta P_{He})} \qquad (23)$$

and for the long path length:

$$R_2 \equiv \frac{T(2)}{T_F(2)} = \frac{1 + P_0 \cdot \tanh(2\eta P_{He})}{1 - F \cdot P_0 \cdot \tanh(2\eta P_{He})} \qquad (24)$$

Subscript "F" identifies the data measured with the spin-flipped polarization. The sign of the product $P_0 \cdot \tanh(\eta P_{He})$ in (23 – 24) can be either positive or negative depending on the signs of $P_0$ and $P_{He}$. So, the ratios $R_1$ and $R_2$ can be either greater or less than unity. We will show later that these two options may bring a different accuracy in the measured neutron polarization. Taking into account that

$$\tanh(2\eta P_{He}) = \frac{2 \cdot \tanh(\eta P_{He})}{1 + \tanh^2(\eta P_{He})} \qquad (25)$$

one can solve (23) and (24) for $\tanh(\eta P_{He})$ and $P_0$.

$$\tanh(\eta P_{He}) = \pm \sqrt{2 \frac{R_1 - 1}{R_2 - 1} \cdot \frac{R_2 F + 1}{R_1 F + 1} - 1} \qquad (26)$$

$$P_0 = \frac{R_1 - 1}{(R_1 F + 1) \cdot \tanh(\eta P_{He})} \qquad (27)$$

Here signs "+" and "-" in (26) indicate the state of the polarization of a $^3$He gas (nuclear spin parallel or antiparallel to the applied magnetic field, respectively). It is worth noting that in the proposed method there is no need to know the individual value of the transmittance in each case since only the ratio of transmittances is involved. Thus, one can write for (23) and (24) the following relations:

$$R_1 \equiv \frac{T(1)}{T_F(1)} = \frac{n_1}{n_{F1}} \qquad (28)$$

$$R_2 \equiv \frac{T(2)}{T_F(2)} = \frac{n_2}{n_{F2}} \qquad (29)$$

with $n_1$ and $n_{F1}$ to be the numbers of neutrons counted with the cell in the position "1" (with direct and reversed polarization of the neutron beam, respectively) and $n_2$ and $n_{F2}$ - numbers of neutrons counted with the cell in the position "2". Thus, formulas (26 - 27) allow extracting a neutron polarization knowing only count rates and flipping efficiency. The gas polarization and cell opacity have not to be known a priori. If a spin-flipper with known efficiency is already integrated in a neutron instrument, then a 2×1 NSF method seems to be a good choice for express analysis of a neutron polarization with required accuracy. This accuracy obviously depends on the errors in the values F, $R_1$ and $R_2$ and will be analyzed below.

## 5. Accuracy of the method

Assuming small errors in measured values one may apply the error propagation method to (26 – 27) and write a standard deviation $\sigma(P_0)$ for the calculated value $P_0$ as

$$\sigma^2(P_0) = A_1^2 \cdot \sigma^2(R_1) + A_2^2 \cdot \sigma^2(R_2) + A_F^2 \cdot \sigma^2(F) \tag{30}$$

where $\sigma(R_{1,2})$ – standard deviation of the measured ratio $R_{1,2}$ and $\sigma(F)$ – standard deviation of the flipping efficiency F. With a view to simplifying the following analysis we write the factors $A_1$, $A_2$ and $A_F$ in (30) as follows:

$$A_1 = \frac{(1 - FP_0 \cdot \tanh(\eta P_{He}))^2}{(1+F) \cdot \tanh(\eta P_{He})} \cdot \left(1 - \frac{1}{\tanh(\eta P_{He}) \cdot \tanh(2\eta P_{He})}\right) \tag{31}$$

$$A_2 = \frac{1}{(1+F) \cdot \tanh(\eta P_{He})} \cdot \left(\frac{1}{\tanh(2\eta P_{He})} - FP_0\right)^2 \tag{32}$$

$$A_F = -\frac{P_0}{(1+F)} \cdot \left(1 + \frac{P_0}{\tanh(2\eta P_{He})}\right) \tag{33}$$

Equations (31 – 33) written in this form can be readily applied to study the effect of NSF parameters on the accuracy of a neutron polarization measurement. The first obvious result is that if $\tanh(\eta P_{He}) \to 1$ then $A_1 \to 0$ while $A_2 \neq 0$ (with reasonable assumption that $FP_0 \neq 1$). It comes from the view of (30) that in this case statistical accuracy is dominated by the numbers measured with the cell in the position "2". It can be understood from the comparison of (23) and (24). Indeed, as the product $\eta P_{He}$ becomes higher, so $\tanh(2\eta P_{He})$ approaches unity much faster against $\tanh(\eta P_{He})$ and hence a NSF in the position "2" first approaches the "ideality" (see text above and ref. [11]). In this case the value of $R_2$ on its own can give $P_0$, provided the value F is known. Nevertheless, by measuring both $R_1$ and $R_2$ one can obtain the analyzing efficiency of the NSF (26) and hence the cell in the position "1" with much higher transmission becomes a well-defined analyzer.

Another important result that can be seen from (31 – 33) concerns the choice of the signs for polarizations $P_0$ and $P_{He}$. Indeed, the absolute values of $A_1$ and $A_2$ become higher if $P_0 \tanh(\eta P_{He}) < 0$ and smaller if $P_0 \tanh(\eta P_{He}) > 0$ while for $A_F$ the reverse situation occurs. This provides a way of controlling the total error by choosing appropriate signs for both polarizations according to which uncertainty dominates in a specific measurement. For reference, the factors $|A_1|$, $|A_2|$ and $|A_F|$ are plotted in Fig.1 ($P_0 \tanh(\eta P_{He}) > 0$) and Fig.2 ($P_0 \tanh(\eta P_{He}) < 0$) as functions of the product $\eta P_{He}$ for the case when F = 0.95 and $P_0$ = 0.95.

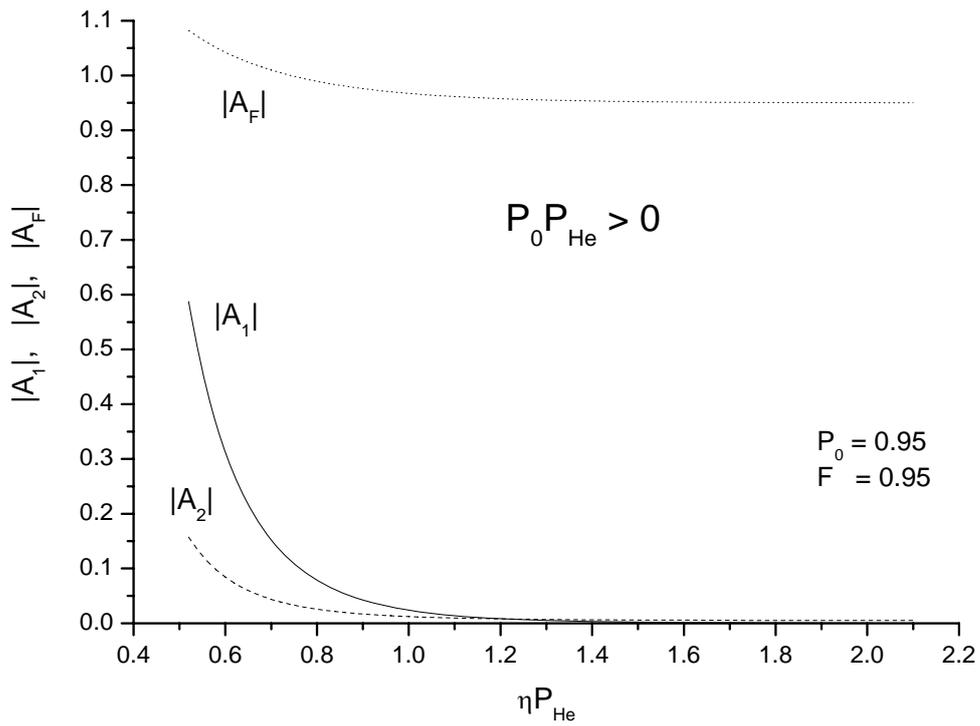

Fig.1 Factors $A_1$, $A_2$ and $A_F$ (see (30 – 33)) vs. product $\eta P_{He}$ at $P_0 P_{He} > 0$.

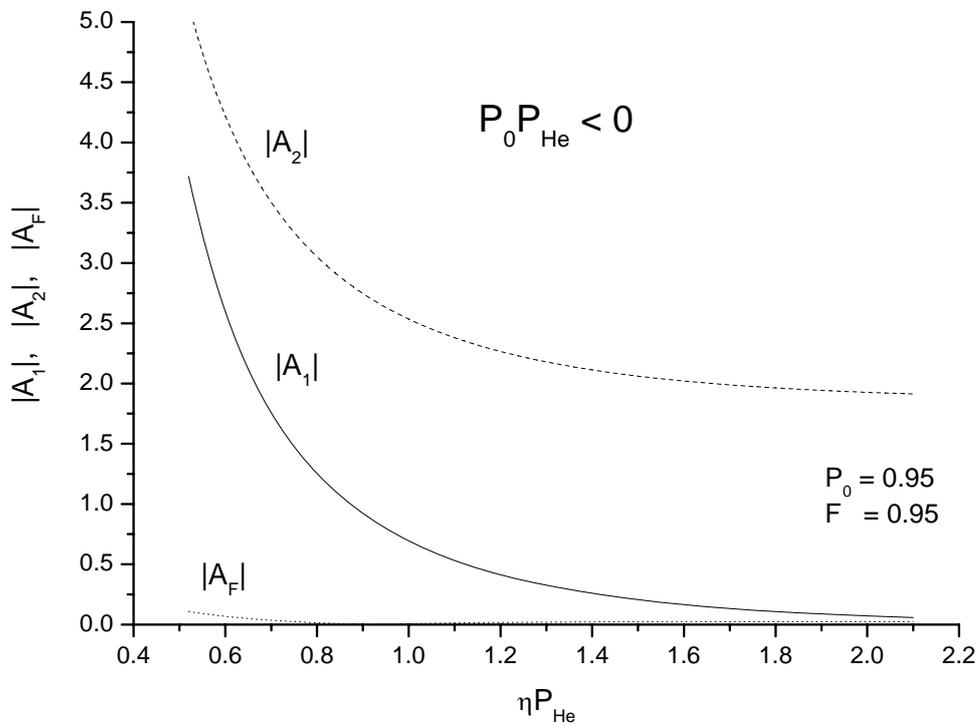

Fig.2 Factors $A_1$, $A_2$ and $A_F$ (see (30 – 33)) vs. product $\eta P_{He}$ at $P_0 P_{He} < 0$.

One can see from Fig.1 that in the case when a neutron beam is polarized parallel to the $^3$He spins the uncertainty in the flipping efficiency becomes governing factor. In principle, the flipping efficiency can be measured with accuracy down to 0.05% [20] and if it is so, then the condition $P_0 \tanh(\eta P_{He}) > 0$ seems to be a good choice. By contrast, if the flipping efficiency is known with poor accuracy then the condition $P_0 \tanh(\eta P_{He}) < 0$ could be preferable. One may expect that even in this case a neutron polarization can be determined with high precision limited only by statistical errors.

Although all the abovementioned results have been obtained for monochromatic neutrons, they can be readily applied to narrow-band neutron beams, which are widely used in experimental setups. Indeed, within a narrow bandwidth the transmittance of a cell can be considered to high accuracy as a linear function of a neutron wavelength $\lambda$ and, assuming a wavelength distribution function $f(\lambda)$ to be symmetrical about the central wavelength $\lambda_0$, one can derive that

$$\int_\lambda T(\lambda) f(\lambda) d\lambda = T(\lambda_0) \cdot \int_\lambda f(\lambda) d\lambda \tag{34}$$

For the flipping ratio it follows then

$$R = \frac{\int_\lambda T(\lambda) f(\lambda) d\lambda}{\int_\lambda T_F(\lambda) f(\lambda) d\lambda} = \frac{T(\lambda_0)}{T_F(\lambda_0)} \tag{35}$$

and hence, a narrow-band neutron beam can be treated as a monochromatic beam with the wavelength $\lambda_0$. If the polarization of a neutron beam with a broad bandwidth has to be measured then a time-of-flight method is required to resolve the beam into its wavelength components. The polarization of each component can be determined through the use of the proposed method and the mean value is calculated then straightforward.

## 6. Nonideal 2×1 NSF cell

From the practical point of view it is also necessary to estimate an error that may arise from the geometrical imperfection of the cell, e.g. when the ratio of the lengths is not exactly equal to 2. Let us assume this ratio to be $2 \cdot (1 + \delta)$ and consider the case when $\delta \ll 1$. Performing straightforward calculations we arrive at the following formula for the related uncertainty in measured polarization

$$\frac{\Delta P_0}{P_0} = B(\eta P_{He}) \cdot \delta \tag{36}$$

where

$$B(\eta P_{He}) = \frac{2\eta P_{He}}{\tanh(\eta P_{He}) \cdot \sinh^2(2\eta P_{He})} \tag{37}$$

The function B($\eta P_{He}$) is displayed in Fig.3.

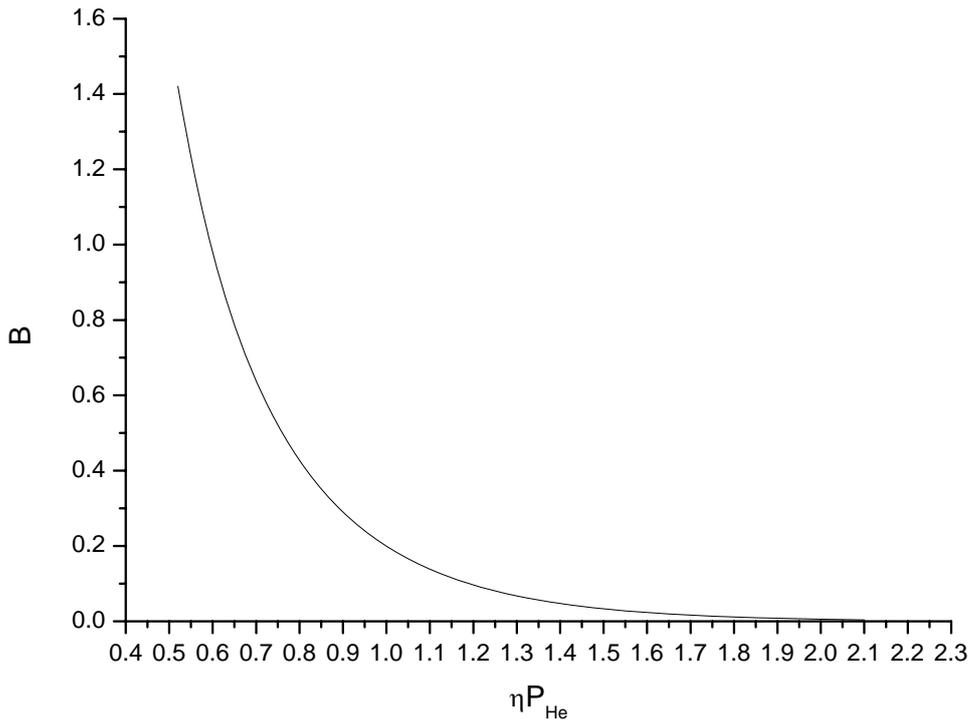

Fig.3  Factor B (see (34 – 35)) vs. product ηP$_{He}$.

From Fig.3 one can see that for $\eta P_{He} \geq 1.2$ the geometrical imperfection of the cell results in measured polarization uncertainty less than 0.1δ. One may expect the geometrical imperfection of the cell can be reduced down to $\delta < 0.01$ thus allowing the error in neutron polarization related to this imperfection to be around 0.001 or less.

## 7. Conclusion

The 2×1 design of a Neutron Spin Filter cell is proposed. In the case when a NSF is used as a polarizer it was shown that the polarization of a passed beam could be determined by measuring transmissions along two different lengths of the same cell only. An application of the 2×1 NSF for an absolute measurement of the polarization of an incident polarized beam is then considered and an extended method is proposed. In this method the transmission of the neutron beam with initial polarization and with reversed one is measured alternately along two different lengths. It was shown that the uncertainties in either $^3$He gas polarization or cell opacity do not affect the accuracy of the measured neutron polarization and this accuracy is mostly limited by statistical errors.

## 8. Acknowledgments

This research project has been supported by the European Commission under the 6th Framework Programme through the Key Action: Strengthening the European Research Area, Research Infrastructures.  Contract n°: RII3-CT-2003-505925